# Symmetry-engineered and electrically tunable in-plane anomalous Hall effect in oxide heterostructures

Kunjie Dai[1,6], Zhen Wang[2,6], Wenfeng Wu[3,4,6], Feng Jin[1], Enda Hua[1], Nan Liu[4], Jingdi Lu[1], Jinfeng Zhang[1], Yuyue Zhao[1], Linda Yang[1], Kai Liu[1], Huan Ye[1], Qiming Lv[1], Zhengguo Liang[1], Ao Wang[1], Dazhi Hou[1,2], Yang Gao[2], Shengchun Shen[1,2], Jing Tao[2], Liang Si[4]*, Wenbin Wu[1,5]*, and Lingfei Wang[1]*

[1] Hefei National Research Center for Physical Sciences at the Microscale, University of Science and Technology of China, Hefei 230026, China.

[2] Department of Physics, University of Science and Technology of China, Hefei 230026, China.

[3] Key Laboratory of Materials Physics, Institute of Solid State Physics, HFIPS, Chinese Academy of Sciences, Hefei 230031, China.

[4] School of Physics, Northwest University, Xi'an 710127, China.

[5] Anhui Key Laboratory of Condensed Matter Physics at Extreme Conditions, High Magnetic Field Laboratory, Chinese Academy of Sciences, Hefei 230031, China.

[6] These authors contribute equally to this work: Kunjie Dai, Zhen Wang, Wenfeng Wu.

* Corresponding author. Email: wanglf@ustc.edu.cn, wuwb@ustc.edu.cn, siliang@nwu.edu.cn.

## Abstract

The family of Hall effects has long served as a premier probe of how symmetry, magnetic order, and topology intertwine in solids. Recently, the in-plane anomalous Hall effect (IP-AHE)—a transverse Hall response driven by in-plane magnetization—has emerged as a distinct member of this family, offering innovative spintronic functionalities and illuminating intricate interplay between mirror-symmetry breaking and in-plane magnetic order. However, practical routes to deterministically and reversibly control IP-AHE remain limited. Here, we establish a symmetry-engineered IP-AHE platform —$CaRuO_3/La_{2/3}Ca_{1/3}MnO_3/CaRuO_3$ heterostructure on $NdGaO_3(110)$— that turns strict mirror-symmetry breaking constraints into effective tuning knobs. IP-AHE in these epitaxial trilayers unambiguously couples to the $CaRuO_3$-buffer-induced mirror-symmetry breaking and faithfully reproduces the ferromagnetic hysteresis. Ionic liquid gating further enables reversible reconfigurations of the symmetry breaking, thereby achieving electrical modulation and ON/OFF switching of IP-AHE. This highly tunable IP-AHE platform opens pathways for exploring nontrivial magnetic order and developing programmable Hall functionalities in planar geometries.

Anomalous Hall effect (AHE) has played a central role over the past century in revealing intriguing magnetic/topological order in solids and enabling a wide range of spintronic functionalities[1-3]. In the conventional AHE geometry, the three vectors, magnetization (*M*), longitudinal electric field (*E*), and transverse Hall current density ($j_{AHE}$), are mutually orthogonal. In magnetic thin films and layered magnets, this orthogonal geometry makes the AHE signal sensitive to the out-of-plane *M* component but insensitive to the in-plane component[4-6]. In practical spintronic devices, this orthogonal constraint restricts Hall readout of magnetic switching in out-of-plane geometries only[3]. Nevertheless, the Onsager relationship does not mandate such strict orthogonality for AHE[7]. Recent theoretical calculations also suggest that breaking the time-reversal- and mirror-symmetries simultaneously may allow an in-plane *M* to generate a transverse Hall response, giving rise to the so-called in-plane AHE (IP-AHE)[8-13]. Subsequent magnetotransport experiments on layered magnets have corroborated this scenario[14-18]. For example, in the $V_2S$-VS superlattice with *C2/m* symmetry, IP-AHE emerges when an in-plane *M*⊥[010] breaks the remaining mirror-symmetry[14]. In the ferromagnetic semimetal $Fe_3Sn_2$ and $Co_3Sn_2S_2$ with $R\bar{3}m$ symmetry, IP-AHE arises from the in-plane *M*⊥[100] axis, which eliminates the three-fold perpendicular mirror planes[16,17]. These advances established the central role of mirror-symmetry breaking in activating IP-AHE. However, due to the strict symmetry requirements, experimental observations of IP-AHE have so far been limited to weakly magnetized two-dimensional materials[14,15], Kagome magnets[16], and magnetic Weyl semimetals[17,18], most of which exhibit weak in-plane magnetic anisotropy (MA). This narrow materials landscape has so far hindered the establishment of a definitive, one-to-one correspondence among mirror-symmetry breaking, IP-AHE, and in-plane ferromagnetism. Consequently, experimentally accessible means for effective control of IP-AHE remain largely undeveloped, posing a major obstacle to its device implementation.

In contrast to the aforementioned layered magnets with nearly invariant structural symmetry, $ABO_3$ perovskite oxide heterostructures can offer highly tunable structural symmetry and magnetism. The lattice symmetries of perovskite oxides are intimately tied to the deformation and collective rotation patterns of the $BO_6$ oxygen octahedra framework[19,20]. Specific oxygen octahedral rotation (OOR) patterns (in Glazer notation), such as $a^0a^0a^0$, $a^-a^-a^-$, and $a^-a^-c^+$, lead to cubic, rhombohedral, and orthorhombic symmetries, respectively[21,22]. By stacking the perovskite oxides into epitaxial heterostructures, their OOR patterns and deformation modes can be flexibly modulated through various factors, such as epitaxial strain, interface structure, proton-intercalation, and off-stoichiometry[19]. Meanwhile, the intricate interplay among spin-orbit coupling (SOC) and electron-electron correlations in these heterostructures gives rise to a rich array of magnetic orders, along with highly tunable MA and AHE[3-5, 23-26]. Furthermore, the strong coupling between the spin, lattice, orbital, and charge degrees of freedom ensures tight links among the structural symmetry, magnetic order, and AHE[27-29]. Collectively, these attributes make perovskite oxide heterostructures an ideal platform for boosting the tunability of IP-AHE.

In this study, we exploit the structural tunability of perovskite oxides to design a symmetry-engineered IP-AHE platform: a $CaRuO_3/La_{2/3}Ca_{1/3}MnO_3/CaRuO_3$ (CRO/LCMO/CRO) epitaxial heterostructure, which converts stringent symmetry breaking constraints into deterministic and effective tuning knobs for IP-AHE. In this ferromagnetic trilayer heterostructure, $CaRuO_3$ buffer-induced mirror-symmetry breaking and uniaxial in-plane magnetic anisotropy cooperatively activate a highly tunable IP-AHE that faithfully reproduces the square-shaped ferromagnetic hysteresis. Ionic liquid gating further introduces reversible protonation in the $CaRuO_3$ buffer to reconfigure the symmetry-breaking *in situ*, enabling wide-range electrical modulation and ON/OFF switching of IP-AHE. This symmetry-tailored, electrically controllable IP-AHE platform can serve as a sensitive probe of nontrivial in-plane spin textures/magnetic orders and open pathways towards programmable Hall functionalities in spintronic devices with planar geometries.

## Observation of IP-AHE in CRO/LCMO/CRO/NGO(110)$_O$ trilayers

We deposit a series of CRO/LCMO/CRO trilayer heterostructures on the (110)$_O$-oriented $NdGaO_3$ [NGO(110)$_O$, where the subscript "O" denotes orthorhombic indices] substrates by pulsed laser deposition (Fig. 1a). The perovskite-structured LCMO, CRO, and NGO share the same orthorhombic *Pbnm* (No. 62) symmetry, arising from the collective in-plane and out-of-plane OOR ($a^-a^-c^+$ in Glazer notation)[19,22]. Reflection high-energy electron diffraction monitoring confirms that both the CRO and LCMO layers grow in a standard layer-by-layer mode (Supplementary Fig. 1), enabling a unit-cell-level control of the CRO and LCMO layer thicknesses [denoted as $t_{CRO}$ and $t_{LC}$, in unit-cells (u.c.)]. High-resolution X-ray diffraction (XRD) $2\theta$-$\omega$ linear scans and atomic force microscopy confirm the high epitaxial quality of the CRO/LCMO/CRO trilayer heterostructures (Supplementary Fig. 2). The XRD reciprocal space mappings (RSMs) further demonstrate that these heterostructures are coherently strained to the NGO(110)$_O$ substrates (Supplementary Fig. 3).

Consistent with the excellent epitaxial quality and structural coherency, the CRO/LCMO/CRO/NGO(110)$_O$ trilayer heterostructure displays strong ferromagnetism with an in-plane uniaxial MA. The temperature-dependent $M$ ($M$-$T$) and longitudinal resistivity ($\rho_{xx}$-$T$) curves of the trilayers reveal clear paramagnetic-insulator to ferromagnetic-metal (FMM) transitions upon cooling (Supplementary Fig. 4). The high Curie temperature ($T_C \geq 250$ K) indicates that the FMM state originates predominantly from the LCMO layer rather than the CRO layer or the CRO/LCMO interface. The increases in $T_C$ and saturated $M$ ($M_s$) with $t_{CRO}$ suggest that CRO layers stabilize the FMM phase in LCMO, likely via interfacial charge transfer from $Ru^{4+}$ to $Mn^{4+}$ cations[30-32]. As depicted in Fig. 1b, we measure the magnetic field-dependent $M$ ($M$-$H$) curves of a representative CRO(16 u.c.)/LCMO(32 u.c.)/CRO(16 u.c.)/NGO(110)$_O$ at 10 K (Figs. 1c-e), with $H$ applied along the [001]$_O$ axis, [$\bar{1}$10]$_O$ axis, and the film normal [⊥(110)$_O$ plane]. The out-of-plane $M$-$H$ curves [$H$⊥(110)$_O$ plane] display negligible $M$ (Fig. 1c), and the in-plane curve measured with $H$//[001]$_O$ displays a slanted hysteresis loop (Fig. 1e), where $M$ saturates at a saturation field $H_S \approx$ 2000 Oe. In sharp contrast, $H$//[$\bar{1}$10]$_O$ yields a square-shaped ferromagnetic hysteresis loop with a small coercive field $H_C \sim 200$ Oe (Fig. 1d), establishing the in-plane magnetic easy-axis along LCMO[$\bar{1}$10]$_O$. Since LCMO intrinsically has weak SOC and negligible magnetocrystalline anisotropy, the in-plane uniaxial MA is governed primarily by demagnetization energy and magnetoelastic contributions arising from NGO(110)$_O$-imposed anisotropic strain[33]. Additionally, the CRO layers, which possess stronger SOC, likely amplify this magnetoelastic coupling and thereby reinforce a robust in-plane uniaxial MA[5,34].

Together with the in-plane uniaxial MA, the CRO/LCMO/CRO/NGO(110)$_O$ trilayers exhibit unique Hall transport. Consistent with the $M$-$H$ measurements, we apply $H$ along the three orthogonal axes, set $E$//[$\bar{1}$10]$_O$ axis, and measure the transverse Hall resistivity ($\rho_{xy}$) along the [001]$_O$ axis (Fig. 1b). We subtract the linear component of the $\rho_{xy}$-$H$ curves by fitting the $H > H_S$ regimes and obtain the $H$-dependent anomalous Hall resistivity ($\rho_{AHE}$-$H$) curves. In the conventional Hall geometry [$H$⊥(110)$_O$], the $\rho_{AHE}$ signal is negligible (Fig. 1f), consistent with the in-plane MA revealed in Figs. 1c-e. When $H$//[$\bar{1}$10]$_O$ (the in-plane easy axis), the $\rho_{AHE}$–$H$ curve surprisingly shows an inverted ferromagnetic hysteresis loop (Fig. 1g), which closely tracks the line shape of the corresponding $M$-$H$ hysteresis loop (Fig. 1d). The $\rho_{AHE}$ signal, proportional to the in-plane $M$ along [$\bar{1}$10]$_O$, strongly suggests the emergence of IP-AHE. By contrast, when $H$//[001]$_O$ (in-plane magnetic hard axis), the $\rho_{AHE}$-$H$ curve (Fig. 1h) displays an unusual hysteresis loop: at $H= 0$, $\rho_{AHE}$ remains finite, close to the value for $\boldsymbol{H}//[\bar{1}10]_O$. Upon increasing $H$, $\rho_{AHE}$ gradually decreases and eventually vanishes near $H_S \approx 2000$ Oe. When $H < H_S$, the in-plane uniaxial MA keeps a sizable $\boldsymbol{M}$ component aligned with the easy axis $[\bar{1}10]_O$ (Figs. 1e, h), thereby sustaining a finite $\rho_{AHE}$. These observations confirm that the $[\bar{1}10]_O$-polarized $\boldsymbol{M}$ component is essential for activating IP-AHE.

## Mirror-symmetry breaking in CRO/LCMO/CRO/NGO(110)$_O$ trilayers

The strong coupling between in-plane uniaxial MA and IP-AHE motivates us to evaluate the structural symmetries of CRO/LCMO/CRO/NGO(110)$_O$ trilayer heterostructures. The XRD RSMs near NGO(420)$_O$, (240)$_O$, (332)$_O$, and (33$\bar{2}$)$_O$ diffractions (Figs. 2a-d) confirm the orthorhombic symmetry of LCMO, CRO, and NGO unit-cells. According to detailed analyses of these RSM results (Supplementary Figs. 5, 6), these oxides have distinct orthorhombicity (defined as $d_{OR} = 1 - a_O/b_O$) but grow in a consistent axis-to-axis epitaxial relationship: CRO[100]$_O$//LCMO[100]$_O$//NGO[100]$_O$ and CRO[010]$_O$//LCMO[010]$_O$//NGO[010]$_O$[35,36]. For CRO and NGO, $d_{OR} > 0$, which is the direct consequence of the strong $a^-a^-c^+$-type OOR[35-37]. By contrast, LCMO shows $d_{OR} < 0$, which can be attributed to the Jahn-Teller distortion of the MnO$_6$ octahedra[38,39].

The heterointerfaces also play a key role in determining the structural symmetry of the LCMO layer in the heterostructure (Supplementary Figs. 7-9). The cross-sectional scanning transmission electron microscopy (STEM) images measured in high-angle annular dark-field (HAADF) mode confirm that both the top and bottom CRO/LCMO heterointerfaces in the trilayer heterostructure are atomically-sharp (Supplementary Fig. 9). Although bulk LCMO is orthorhombic (space group: *Pbnm*), the top and bottom CRO/LCMO heterointerfaces will further eliminate specific mirror planes (M$_{(010)}$ and M$_{(100)}$) and rotation axes (2$_{[100]}$ and 2$_{[010]}$), thus lowering the structural symmetry to monoclinic (space group: *P*2$_1$/*m*, see detailed analyses in Supplementary Fig. 7, Supplementary Tables 1 and 2).

The symmetry lowering in our CRO/LCMO/CRO trilayer can be visualized by converting the orthorhombic unit-cell into the pseudo-cubic unit-cell (denoted by the subscript "p", see Supplementary Fig. 8 for details). In these pseudo-cubic unit-cells, the $c_p$ axis tilts towards the $a_p$ axis while remaining perpendicular to the $b_p$ axis ($\alpha = 90°$ but $\beta \neq 90°$), yielding a monoclinic-like distortion (as depicted in Fig. 2e). The layer-averaged tilting angle $\beta$ can be directly calculated from the orthorhombicity $d_{OR}$[35,40,41]: for CRO and NGO with $d_{OR} > 0$, $\beta_{CR} < \beta_{NG} < 90°$, but for LCMO with $d_{OR} < 0$, $β_{LC} > 90°$ (Fig. 2e). The uniform but opposite monoclinic-like tilting in the CRO and LCMO unit-cells can be further confirmed by geometric phase analysis (GPA) on cross-sectional HAADF-STEM images (Figs. 2f-h and Supplementary Section 3) and even visible directly in an atomic-scale STEM-HAADF image (Fig. 2i). High-resolution STEM images in annular bright-field (ABF) mode further reveal the $BO_6$ octahedral coupling across the LCMO/CRO interfaces (Supplementary Fig. 11), which enables the axis-to-axis epitaxial relationship and the opposite monoclinic-like tilting distortions in CRO and LCMO. Moreover, the stronger OOR in CRO can propagate into LCMO by several u.c., enhancing the Jahn-Teller distortion and thus enlarging the $d_{OR}$[42-44]. Consequently, the LCMO layer in the trilayer heterostructure shows a stronger monoclinic-like distortion than both the bulk LCMO and the single-layer LCMO/NGO(110)$_O$ film (Supplementary Fig. 12).

The RSM and STEM characterizations of the CRO/LCMO/CRO heterostructures suggest that the $MnO_6$ octahedral rotation/deformation and atomically-sharp interfaces cooperatively lower the structural symmetry of LCMO layers to monoclinic ($P2_1/m$). In this monoclinic-distorted LCMO unit-cell, only one vertical mirror plane (⊥[001]$_O$) and a twofold rotation axis (along [001]$_O$) remain. Accordingly, the monoclinic tilting angle $β_{LC}$ can be seen as a reliable parameter for evaluating the degree of mirror-symmetry breaking in the LCMO layer. Notably, the remaining symmetry operations define a $C_{2h}$ point group, which allows $M$ along the [$\bar{1}$10]$_O$ axis (the in-plane easy axis) further breaks the remaining mirror-symmetry (Fig. 3a) for activating IP-AHE[8,13].

**Symmetry-engineered IP-AHE**

Having established the microscopic structural symmetry of the CRO/LCMO/CRO trilayer heterostructures, we now turn to clarifying the origin of IP-AHE from the viewpoint of mirror-symmetry breaking. The relationship between intrinsic AHE current density $\boldsymbol{j}_{AHE}$, net Berry curvature ($\bar{\boldsymbol{\Omega}}$) vector, and the applied $\boldsymbol{E}$ is described by the following equation[2,3,8]:

$$\boldsymbol{j}_{AHE} = \frac{e^2}{\hbar} \bar{\boldsymbol{\Omega}} \times \boldsymbol{E} \qquad (1)$$

where the $\bar{\Omega} = \int [d\boldsymbol{k}]\, \Omega(\boldsymbol{k})\, f_0$, with $\Omega(\boldsymbol{k})$ the Berry curvature at crystal momentum $\boldsymbol{k}$ and $f_0$ the Fermi distribution function. In the conventional AHE case, $\bar{\boldsymbol{\Omega}}$ vector is parallel to the out-of-plane $\boldsymbol{M}$ vector, enforcing the orthogonal relationship $\boldsymbol{E} \perp \boldsymbol{M} \perp \boldsymbol{j}_{AHE}$. As mentioned above, the monoclinic-distorted LCMO layer satisfies the mirror-symmetry breaking restriction for activating IP-AHE, which allows in-plane $\boldsymbol{M}$//[$\bar{1}$10]$_O$ generates a non-zero $\bar{\boldsymbol{\Omega}}$ along film normal ($\bar{\Omega}_z$). The three vectors ($j_{AHE}$, $E$, and $M$) then become coplanar, thus producing a pronounced IP-AHE[8, 10, 13]. Conversely, when $M$//[001]$_O$, the (001)$_O$ mirror plane is preserved, thus suppressing $\bar{\Omega}_z$ and the IP-AHE (Fig. 1h).

To verify the key role of mirror-symmetry breaking-activated $\bar{\Omega}_z$ in inducing the IP-AHE, we track the evolution of IP-AHE with the directions of $E$ and $H$. First, with fixed $E$//[$\bar{1}$10]$_O$, we rotate the $H$ vector in-plane (the angle between $H$ and $E$ is defined as $\theta$, Fig. 3a). As the $H$ vector rotates from the easy axis ($H$//[$\bar{1}$10]$_O$, $\theta = 0°$) to the hard axis ($H$//[001]$_O$, $\theta = 90°$), the saturated $\rho_{AHE}$ at $H$ = 2000 Oe decreases gradually to zero (Figs. 3b, c). This is consistent with our speculation: only the in-plane $M$//[$\bar{1}$10]$_O$ produces the IP-AHE. The finite $\rho_{AHE}$ observed for $0° < \theta \leq 90°$ should arise from the residual $M$ component pinned to the [$\bar{1}$10]$_O$ easy axis (Supplementary Figs. 13, 14). As $\theta$ increases over 90°, the saturated $\rho_{AHE}$ reverses sign due to a 180° reversal of the $M$ component pinned to the [$\bar{1}$10]$_O$ axis. The resultant cosine-shaped $\rho_{AHE}$-$\theta$ curve (Fig. 3c and Supplementary Figs. 13, 14) corroborates the coplanar relationship between $E$, $M$, and $j_{AHE}$, revealing the dipolar structure of the $\bar{\Omega}$ in spin-order space[13]. Second, we characterize the evolution of $\rho_{AHE}$ by rotating $H$ within the cross-sectional (001)$_O$ plane, and confirm that the IP-AHE is not induced by the out-of-plane component of misaligned in-plane $M$ (Supplementary Figs. 15, 16). Third, with fixed $H$//[$\bar{1}$10]$_O$, we rotate the $E$ in-plane (the angle between $E$ and [$\bar{1}$10]$_O$ axis is defined as $\varphi$, Fig. 3a) by fabricating a series of Hall bars along different orientations (Supplementary Fig. 17). Both the $\rho_{AHE}$-$H$ hysteresis loops (Fig. 3d) and saturated $\rho_{AHE}$ values (Fig. 3e) are independent of the $\varphi$. These results are in sharp contrast to the planar Hall effect but consistent with Eq. 1: a nonzero $\bar{\Omega}_z$-induced $j_{AHE}$ should be insensitive to the in-plane orientation of $E$.

Guided by the above results and analyses, we infer that IP-AHE in the CRO/LCMO/CRO trilayers could be effectively tuned by modulating the degree of mirror-symmetry breaking in LCMO. As established above, the strongly coupled OOR patterns at the CRO/LCMO heterointerfaces could make $t_{CRO}$ a potential tuning knob for modulating the structural symmetry and IP-AHE of the LCMO layer. To verify this tuning scenario, we first calculate the in-plane anomalous Hall conductivity as $\sigma_{AHE} = \rho_{AHE}/(\rho_{xx})^2$ for parameterizing the IP-AHE, thereby eliminating the influence of $t_{CRO}$-dependent $\rho_{xx}$. Figs. 4a-e show the representative $\sigma_{AHE}$-$H$ curves ($H$//[$\bar{1}$10]$_O$) at 10 K, measured from a series of CRO/LCMO/CRO trilayer heterostructures with unequal $t_{CRO}$. The zero-field $\sigma_{AHE}$ values display a non-monotonic $t_{CRO}$-dependence (Fig. 4f). As $t_{CRO}$ increases, $\sigma_{AHE}$ initially increases rapidly, peaks at $t_{CRO}$ = 16 u.c., and then it decreases gradually. Owing to the strong uniaxial in-plane MA and associated square-like $\sigma_{AHE}$-$H$ hysteresis loops, the optimal $\sigma_{AHE}$ at zero-field is up to 103.7 mS·cm$^{-1}$, close to the saturated value (106.3 mS·cm$^{-1}$). Such a sizable zero-field IP-AHE was rarely observed in layered magnets with weak in-plane MA[14-18]. More strikingly, the $\beta_{LC}$ − 90° value (Fig. 4g), signifying the magnitude of monoclinic-like distortion in LCMO (Supplementary Figs. 3, 6, and Supplementary Table 3), follows the same non-monotonic $t_{CRO}$-dependence as the $\sigma_{AHE}$ (Fig. 4f). Because the NGO(110)$_O$ substrate has much weaker OOR than bulk CRO, the interfacial octahedral coupling at the CRO/NGO(110)$_O$ interface is expected to suppress the OOR in the ultrathin CRO buffer. Consequently, as $t_{CRO}$ increases from 0 to 16 u.c., the OOR amplitude in CRO buffer rapidly approaches its bulk value, which increases the $\beta_{LC}$ value, enlarges the mirror-symmetry breaking, and thus increases the $\sigma_{AHE}$. Nevertheless, further increasing $t_{CRO}$ could lead to considerable strain-relaxation and formation of $a/b$ twin domains, thus reducing the $\beta_{LC}$ and $\sigma_{AHE}$.

The IP-AHE of CRO/LCMO/CRO trilayer also shows strong $T$-dependence. As shown in Fig. 4h and Supplementary Fig. 18, nonzero $\sigma_{AHE}$ persists up to 200 K. Upon cooling, all of the $\sigma_{AHE}$-$T$ curves exhibit a clear $\sigma_{AHE}$ increment at $T$ < 80 K, consistent with the reduction of $\rho_{xx}$. For the trilayer samples with relatively weak IP-AHE ($t_{CRO} \geq 32$ u.c. or $t_{CRO} \leq 8$ u.c.), $\sigma_{AHE}$ reverses sign from negative to positive near 100 K upon warming (inset of Fig. 4h). Such a sign-reversal behavior is similar to the intrinsic AHE of SRO thin films, which is sensitive to the band topology[4,23].

The above results in Figs. 3 and 4 strongly indicate that the IP-AHE in CRO/LCMO/CRO trilayers originates from the mirror-symmetry-breaking-activated $\bar{\Omega}_z$ of the LCMO layer. On this basis, we perform first-principles density-functional-theory (DFT) calculations to further elucidate the link between $\bar{\Omega}_z$, structural symmetry, and IP-AHE in LCMO. Considering that the $\beta_{LC}$ serves as a reliable control parameter of the mirror-symmetry breaking and thus the IP-AHE, we construct four different (110)$_O$-oriented LCMO 2×2×2 supercells: one high-symmetry cubic supercell without any OOR and three orthorhombic supercells with $a^-a^-c^+$-type OOR and Jahn-Teller distortion. The variable amplitudes of OOR and Jahn-Teller distortions in the orthorhombic supercells give rise to variable $\beta_{LC}$ = 90.0°, 90.08°, and 90.16° (Supplementary Fig. 19). The Wannier bands show excellent agreement with the DFT results (Supplementary Fig. 20), validating the following AHE and Berry curvature calculations. After imposing $M$ along the $[\bar{1}10]_O$ axis, DFT-calculated $\sigma_{AHE}$ ($\sigma_{DFT}$) vanishes in the cubic supercell but remains finite in all the orthorhombic supercells, even for the one with $\beta_{LC}$ = 90.0° (Figs. 4i, j, and Supplementary Fig. 21). The magnitude of $\sigma_{DFT}$ increases monotonically with $\beta_{LC}$, consistent with the experimental trend. The stark contrasts of $\sigma_{DFT}$ between the cubic and orthorhombic supercells strongly suggest that the MnO$_6$ octahedral rotation/deformation and the associated mirror-symmetry breaking dominate the emergence and tunability of IP-AHE. In the practical CRO/LCMO/CRO/NGO(110)$_O$ trilayer heterostructures, the OOR and Jahn-Teller distortion in LCMO are strongly coupled with the monoclinic-like tilt, making $\beta_{LC}$ the experimental tuning knob of IP-AHE. Following the *H*-direction-dependent IP-AHE shown in Fig. 3a, we also calculate the evolution of non-zero $\sigma_{DFT}$ with the orientation of in-plane *M* vector (Fig. 4k and Supplementary Fig. 22). These results again confirm the critical role of the uniaxial *M*//$[\bar{1}10]_O$ in activating $\bar{\Omega}_z$ and inducing IP-AHE.

**Electrical control of IP-AHE via ionic-liquid gating**

After establishing the OOR-mediated symmetry-engineering of $\bar{\Omega}_z$ in the CRO/LCMO/CRO trilayers, we employ an ionic liquid gating (ILG) approach (see Methods for procedural details) to achieve in situ electrical modulation of IP-AHE. As schematically illustrated in Fig. 5a, a positive gate voltage ($V_g$) electrolyzes residual water in the ionic liquid and then triggers proton intercalation into the CRO/LCMO/CRO trilayers[45]. Previous reports have revealed that such a protonation process can induce both electron doping and structural transformations in perovskite oxide heterostructures[45, 46]. Accordingly, ILG is expected to effectively tune the symmetry-coupled IP-AHE. As shown in Figs. 5b, c, increasing $V_g$ from 0 to 1.0 V leads to negligible changes in the $\sigma_{AHE}$-*H* curves. When $V_g$ exceeds 1.1 V, $\sigma_{AHE}$ starts to decrease progressively and vanishes at $V_g$ = 1.4 V. In contrast to this effective suppression of IP-AHE, the $H_C$ values remain nearly unchanged during protonation, suggesting a nearly invariant macroscopic ferromagnetism. According to previous literature, ILG-driven protonation in ruthenates occurs at a relatively low threshold $V_g \leq$ 1.5 V and completes within a narrow $V_g$ range (~0.5 V)[47, 48]. In contrast, the protonation in manganites typically occurs at a much higher threshold $V_g$ > 2.5 V, and continuously evolves over a wider $V_g$ range (>1.0 V)[49]. Therefore, during the ILG of our CRO/LCMO/CRO trilayer, the sharp changes in $\sigma_{AHE}$ at 1.1 V $\leq V_g \leq$ 1.4 V should coincide with the protonation occurring first in the CRO layers, forming an H$_x$CaRuO$_3$ phase[48]. Furthermore, by alternatively applying positive (+1.4 V) and negative (−1.0 V) $V_g$ to drive structural transformations between CRO and H$_x$CaRuO$_3$, we can reversibly switch the IP-AHE ON and OFF (Supplementary Fig. 23).

Consistent with the evolutions of IP-AHE during ILG, the XRD 2θ-ω linear scans measured in the small $V_g$ range (0~1.5 V) display a clear CRO(440)$_O$ peak shift towards lower Bragg angles, while the LCMO(440)$_O$ peak position remains nearly unchanged (Supplementary Fig. 24). DFT calculations further indicate that the protonation in CRO effectively reduces the OOR in H$_x$CaRuO$_3$[48], especially the out-of-plane tilting [Fig. 5b and Supplementary Figs. 25 and 26], thereby lowering the OOR in the adjacent LCMO layer through interfacial octahedral coupling (Supplementary Fig. 27). Consequently, the CRO/LCMO/CRO trilayer exhibits effective and reversible suppressions of IP-AHE with minimal change in ferromagnetism.

## Conclusion

we established a ferromagnetic CRO/LCMO/CRO/NGO(110)$_O$ heterostructure as a distinct symmetry-engineered and electrically tunable platform for IP-AHE. In this trilayer epitaxial system, the corner-sharing octahedral framework and atomically sharp CRO/LCMO heterointerfaces, together with the uniaxially-aligned in-plane ***M***, cooperatively break the vertical mirror-symmetries in the LCMO layer and activate non-zero $\bar{\Omega}_z$. The resulting IP-AHE faithfully tracks the square-shaped ferromagnetic hysteresis loops, and persists at zero field up to 200 K. By adjusting the $t_{CRO}$, we deliberately reconfigure the degree of mirror-symmetry breaking and establish its intimate correlation with $\sigma_{AHE}$. By introducing protonation via ILG, we further realize a wide-range electrical modulation and reversible ON/OFF switching of IP-AHE.

This symmetry engineering strategy for IP-AHE should be broadly applicable across magnetic heterostructures and layered magnets with rich lattice degrees of freedom. Fundamentally, the symmetry-tailored IP-AHE offers a powerful fingerprint for uncovering intimate links between quantum geometry and nontrivial magnetic orders/spin textures. Practically, the electrically tunable IP-AHE enables planar-geometry-compatible Hall readout of spin transport and magnetic switching, opening an avenue toward programmable Hall functionalities and innovative spintronic and magnetoelectric device architectures.

*Letters* **105**, 227203 (2010).

43. Borisevich, A. Y. *et al.* Suppression of Octahedral Tilts and Associated Changes in Electronic Properties at Epitaxial Oxide Heterostructure Interfaces. *Physical Review Letters* **105**, 087204 (2010).

44. Kan, D. *et al.* Tuning magnetic anisotropy by interfacially engineering the oxygen coordination environment in a transition metal oxide. *Nature Materials* **15**, 432-437 (2016).

45. Lu, N. *et al.* Electric-field control of tri-state phase transformation with a selective dual-ion switch. *Nature* **546**, 124-128 (2017).

46. Guan, Y., Han, H., Li, F., Li, G. & Parkin, S. S. P. Ionic Gating for Tuning Electronic and Magnetic Properties. *Annual Review of Materials Research* **53**, 25-51 (2023).

47. Li, Z. *et al.* Reversible manipulation of the magnetic state in $SrRuO_3$ through electric-field controlled proton evolution. *Nature Communications* **11**, 184 (2020).

48. Shen, S. *et al.* Emergent Ferromagnetism with Fermi-Liquid Behavior in Proton Intercalated $CaRuO_3$. *Physical Review X* **11**, 021018 (2021).

49. Cui, B. *et al.* Reversible Ferromagnetic Phase Transition in Electrode-Gated Manganites. *Advanced Functional Materials* **24**, 7233-7240 (2014).
14

## Methods

**Thin film growth**

We grow epitaxial CaRuO$_3$($t_{CRO}$)/La$_{2/3}$Ca$_{1/3}$MnO$_3$(32 u.c.)/CaRuO$_3$($t_{CRO}$) (CRO/LCMO/CRO) trilayer heterostructures on NdGaO$_3$(110)$_O$ [NGO(110)$_O$] substrates by pulsed laser deposition (KrF excimer laser, $\lambda$ = 248 nm). Before growth, we anneal the NGO(110)$_O$ substrates at 1000 °C for 4 h to obtain single-unit-cell-height surface terraces. During the growth of CRO and LCMO layers, we set the substrate temperature to 700 °C and the laser fluence to 1.8 J cm$^{-2}$. We set the oxygen partial pressure ($P_{O2}$) to 7.5 Pa for CRO and 30 Pa for LCMO. We monitor and optimize the growth rate and mode using reflection high-energy electron diffraction (RHEED). After deposition, we anneal the heterostructures in situ for 15 min under the growth conditions and then cool them down to room temperature in an oxygen atmosphere of $P_{O2}$ = 2000 Pa.

**Structural, Magnetic, and Electrical Characterizations**

We perform the structural characterizations of the trilayer heterostructures using a high-resolution X-ray diffractometer (PANalytical Empyrean, Cu K$\alpha_1$ radiation) in 2$\theta$–$\omega$ scan and reciprocal-space-mapping (RSM) modes. We prepare cross-sectional scanning transmission electron microscopy (STEM) specimens by focused ion beam (Ga$^+$) milling to a thickness of ~30 nm. We perform cross-sectional STEM on a Thermo Scientific Themis Z microscope operated at 300 keV and equipped with a spherical aberration corrector for the condenser lens. We acquire atomic-resolution STEM images using a 25 mrad convergent semi-angle, with collection angles of 41-200 mrad for high-angle annular dark-field (HAADF) and 13-25 mrad for annular bright-field (ABF) imaging. We measure the magnetic properties of the trilayer heterostructures using a vibrating sample magnetometer (VSM-SQUID). We measure temperature- and field-dependent $\rho_{xx}$ and $\rho_{xy}$ using a custom-built physical property measurement system equipped with a rotational sample stage and a 7 T superconducting magnet. To perform the Hall measurements, we pattern the heterostructures into six-probe Hall bars using photolithography.

**Density-Functional Theory Calculations**

We perform density-functional theory (DFT) calculations to investigate the structural, electronic, and magnetic properties of La$_{2/3}$Ca$_{1/3}$MnO$_3$ using the Vienna *ab initio* Simulation Package (VASP)[50, 51], with the Perdew–Burke–Ernzerhof (PBE) generalized gradient approximation (GGA)[52] for exchange–correlation. To model Ca substitution in LaMnO$_3$, we employ the virtual crystal approximation (VCA)[53]. We construct a 2×2×2 supercell (based on the pseudo-cubic unit-cell of La$_{2/3}$Ca$_{1/3}$MnO$_3$) by reorienting the tetragonal unit-cell so that the new *c*-axis aligns with the [110]$_O$ direction of the original lattice, which yields a $\sqrt{2} \times \sqrt{2} \times 1$ expansion of the original tetragonal cell.

We use a plane-wave cutoff energy of 550 eV and fully relax the structure until the residual forces on all atoms are below 0.01 eV/Å. After structural optimization, we calculate the electronic band structure using the WIEN2k package[54,55] with a 7×7×7 Monkhorst–Pack k-point mesh. To account for strong correlation effects and spin–orbit coupling (SOC) in the Mn 3d orbitals, we perform DFT + U + SOC calculations with $U = 8.0$ eV and $J = 2.0$ eV[56]. We project the Mn 3d states onto maximally localized Wannier functions (MLWFs)[57] using Wannier90 code[58] through the Wien2Wannier interface[59]. We then use the resulting tight-binding Hamiltonian to compute the intrinsic anomalous Hall conductivity (AHC) with Wannier90[58] by integrating the Berry curvature over the occupied states across the Brillouin zone.

**Ionic Liquid Gating (ILG)**

Before the in situ XRD and magnetotransport measurements of ILG-induced structural changes, we deposit Pt electrodes along the film edges by DC ion sputtering as bottom contacts, and then we use a thin Pt plate as the top electrode. We then place a droplet of commercial ionic liquid (DEME–TFSI$^{+-}$) to cover both the film surface and the Pt plate. We apply the bias using a Keithley 2450: we hold the potential between the bottom and top electrodes at 0 V and then ramp it to the target value while continuously recording XRD spectra and Hall signals.

**Data availability**

All data relevant to the conclusions of this study are available from

**Acknowledgments**

This work is supported by the National Key Research and Development Program of China (Grant No. 2023YFA1406404), the National Natural Science Foundation of China (Grant Nos. 52521006, 52572144, 12504152, 12374094, 12304153, 12422407, and 12304035), Quantum Science and Technology-National Science and Technology Major Project (Grant No. 2024ZD0301300), CAS Project for Young Scientists in Basic Research (No. YSBR-084), Anhui Provincial Natural Science Foundation (Grant No. 2308085MA15). The computational resource is provided by the National Supercomputing Center (Xi'an) in Northwest University. Synchrotron X-ray diffraction characterizations were supported by BL02U2 at the Shanghai Synchrotron Radiation Facility. The device fabrication was partially carried out at the Center for Micro and Nanoscale Research and Fabrication, University of Science and Technology of China. And the electrical characterizations were carried out at Instruments Center for Physical Science, University of Science and Technology of China.

**Author contributions**

L.W. and W.W. conceived the idea and supervised the project. K.D. completed the theoretical analysis of the symmetry of IP-AHE under the guidance of L.W., W.W. and Y.G.. K.D., Z.W., and Wf.W. contributed equally to this work. K.D. prepared the samples with the help of F.J., E.H., J.L., J.Z, Y.Z. K.D. performed the magnetotransport and electrical transport characterizations with the help of K.L., H.Y., Q.L., Z.L., A.W. K.D. regulated the sample with ILG with the help of S.S. and L.Y. Z.W. supervised the STEM measurements and analyzed the data. Wf.W. and N. L. provided the DFT calculations under the guidance of L.S., L.W., W.W. and K.D. wrote the manuscript, with contributions from all authors.

**Competing interests**

The authors declare no competing interests.

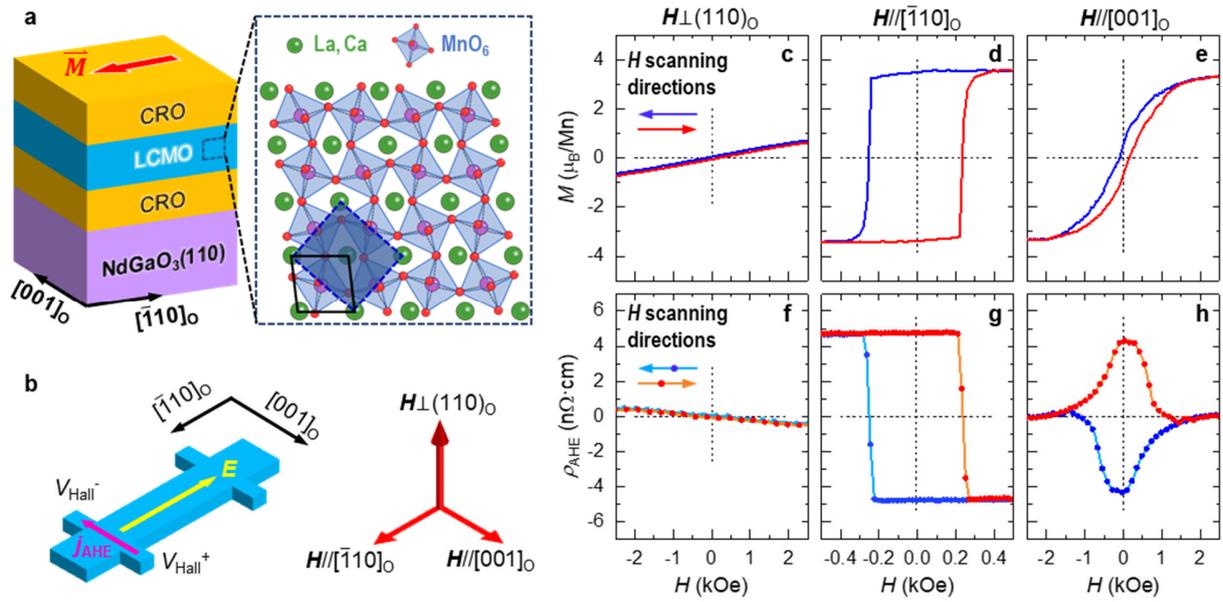

**Fig. 1| Magnetic anisotropy and anomalous Hall effect in CaRuO$_3$/La$_{2/3}$Ca$_{1/3}$MnO$_3$/CaRuO$_3$ trilayer heterostructure.** (**a**) Stacking sequence of CaRuO$_3$/La$_{2/3}$Ca$_{1/3}$MnO$_3$/CaRuO$_3$ (CRO/LCMO/CRO) trilayer heterostructure grown on NdGaO$_3$(110)$_O$ [NGO(110)$_O$] substrate. The inset shows the schematic atomic structure of the LCMO unit-cell, viewed along the [001]$_O$ (in orthorhombic index) axis. The pseudo-cubic and orthorhombic unit-cells are labeled by solid and dashed boxes, respectively. (**b**) Schematic illustration of the traditional (out-of-plane) and in-plane anomalous Hall effect (IP-AHE) measurements. The electric field (***E***) is applied along the [$\bar{1}$10]$_O$ axis and the magnetic field (***H***) is applied along three orthogonal axes: [001]$_O$, [$\bar{1}$10]$_O$, and the film normal [⊥(110)$_O$ plane]. The Hall voltage ($V_{Hall}$) and transverse Hall current density (***j***$_{AHE}$) corresponding to positive anomalous Hall resistivity ($\rho_{AHE}$) are also marked in (b). (**c-h**) *H*-dependent magnetization (*M-H*) curves (c-e) and $\rho_{AHE}$-*H* curves (f-h) measured from the CRO(16 u.c.)/LCMO(32 u.c.)/CRO(16 u.c.)/NGO(110)$_O$ heterostructure at 10 K, with ***H***⊥(110)$_O$ plane (c,f), ***H***//[$\bar{1}$10]$_O$ (d,g), and ***H***//[001]$_O$ (e,h).

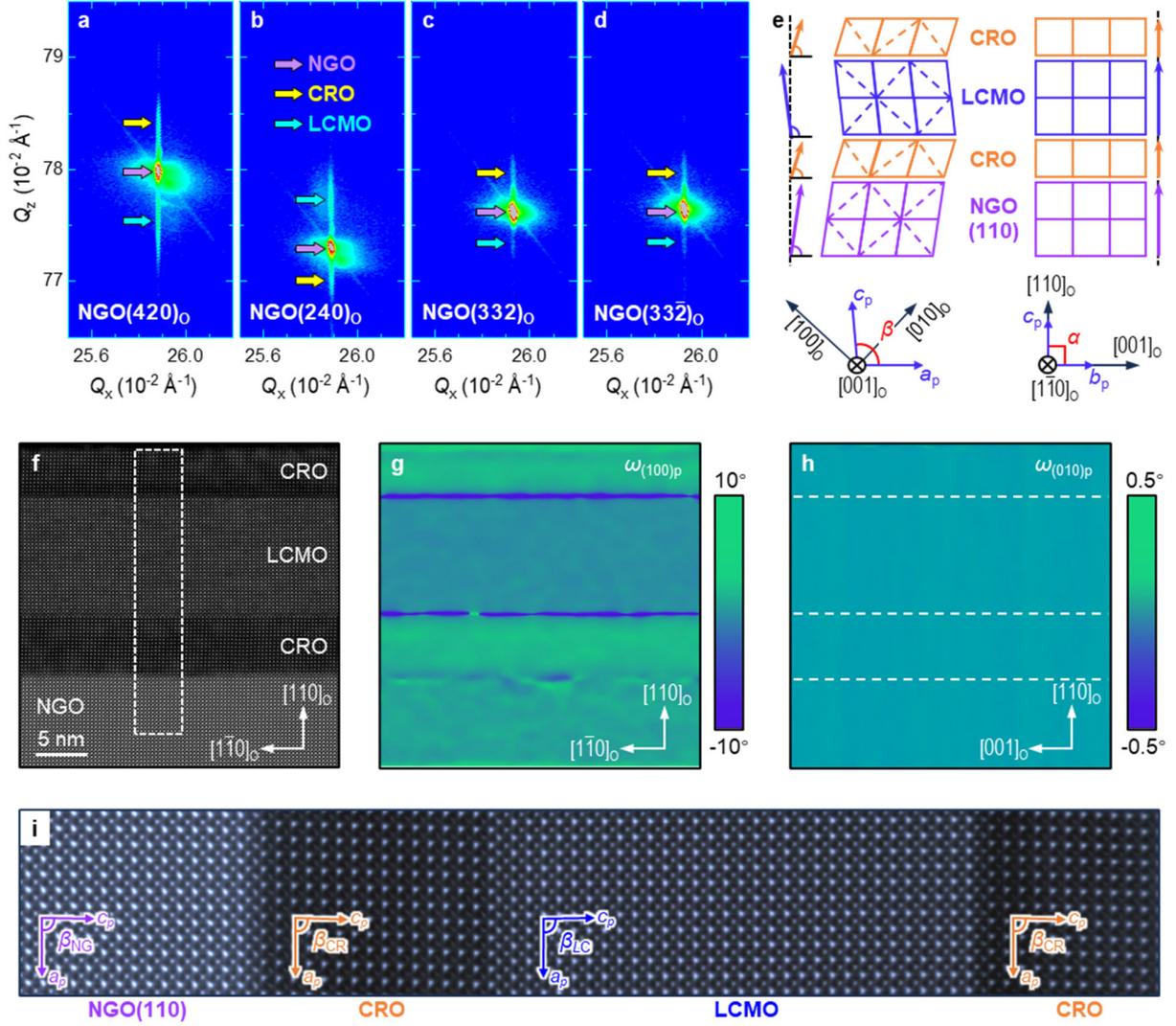

**Fig. 2| Structural characterizations on the CRO/LCMO/CRO trilayer heterostructure.** (**a-d**) Reciprocal space mappings measured from the CRO(16 u.c.)/LCMO(32 u.c.)/CRO(16 u.c.)/NGO(110)$_O$ sample near the NGO(420), (240), (332), and (33$\bar{2}$) diffractions. The diffractions from CRO, LCMO, and NGO are marked by solid arrows in orange, blue, and purple, respectively. (**e**) Schematics of the monoclinic-like tilting of pseudo-cubic CRO, LCMO, and NGO unit-cells, viewed along [001]$_O$ and [$\bar{1}$10]$_O$ axes. The angle between (110)$_O$ and (1$\bar{1}$0)$_O$ planes is labeled as $\beta$. Variations of $\beta$ in CRO, LCMO, and NGO are marked by solid arrows. The angle between (110)$_O$ and (001)$_O$ planes is labeled as $\alpha$. For all of the oxide layers, $\alpha$ = 90°. (**f**) Large-scale HAADF-STEM images of the CRO(16 u.c.)/LCMO(32 u.c.)/CRO(16 u.c.)/NGO(110)$_O$ sample, viewed along the [001]$_O$ zone axis. (**g,h**) Geometric phase analysis of the ($\bar{1}$10)$_O$ [(100)$_p$] plane tilting angle [defined as $\omega_{(100)p}$] (g) and the (001)$_O$ [(010)$_p$] plane tilting angle [defined as $\omega_{(010)p}$] (h). (**i**) A zoom-in HAADF-STEM image measured from the region marked by a white box in (**f**). This high-resolution image can clearly resolve the $\beta$ angles of LCMO, CRO, and NGO.

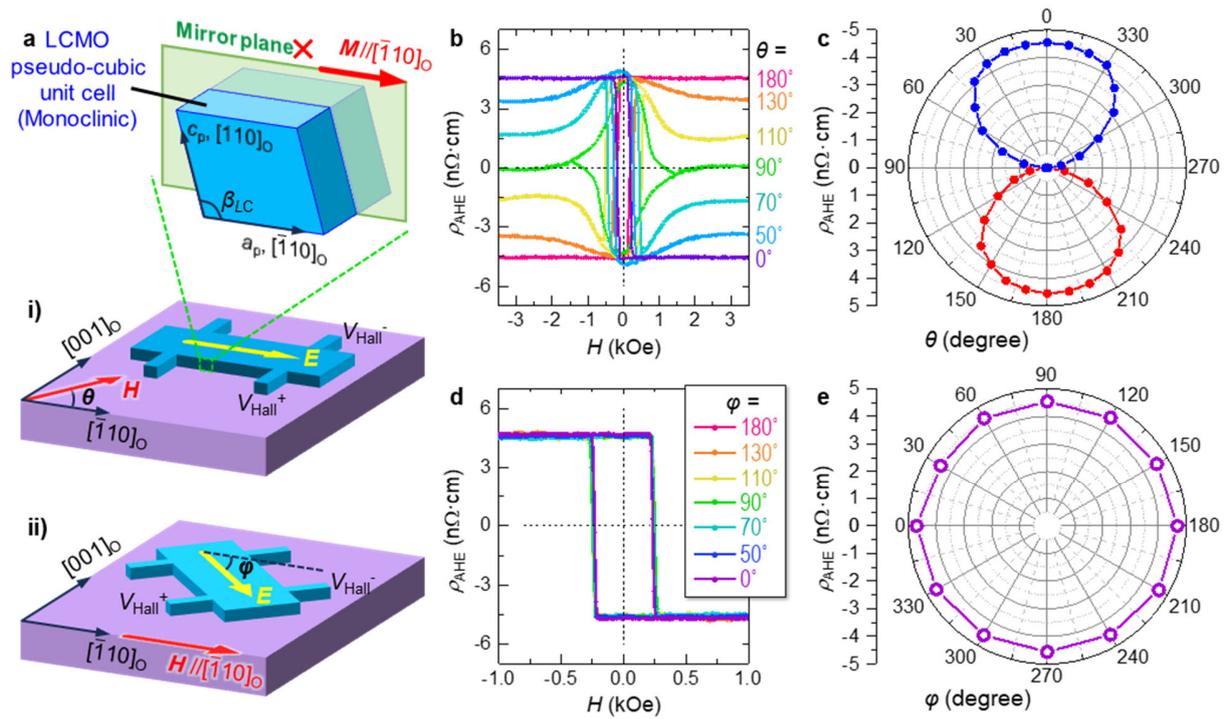

**Fig. 3| IP-AHE of CRO/LCMO/CRO trilayer heterostructure measured at various $H$ and $E$ orientations.** (**a**) Top panel: schematic pseudo-cubic LCMO unit-cell with monoclinic-like distortion. The light green rectangle indicates the only one mirror-symmetry left in LCMO, which can be broken by the $M$ vector parallel to $[\bar{1}10]_O$ axis. Middle and bottom panel: schematic illustrations of IP-AHE measurements with two geometries: i) we fix the applied $E$ direction along LCMO$[\bar{1}10]_O$ axis ($a_p$-axis), and rotate $H$ vector in the LCMO(110)$_O$ plane. The angle between $H$ and LCMO$[\bar{1}10]_O$ axis is defined as $\theta$. ii) We fix the $H$ vector along LCMO$[\bar{1}10]_O$ axis, and rotate the $E$ in the LCMO(110)$_O$ plane. The angle between $E$ and LCMO$[\bar{1}10]_O$ axis is defined as $\varphi$. (**b**) $\rho_{AHE}$-$H$ curves measured in geometry (i) with various $\theta$ values. (**c**) Polar-plot of saturated $\rho_{AHE}$ ($\rho_{AHE}$ at $H = 1000$ Oe) versus $\theta$. (**d**) $\rho_{AHE}$-$H$ curves measured with geometry (ii) for various $\omega$ values. (**e**) Polar-plot of saturated $\rho_{AHE}$ versus $\varphi$.

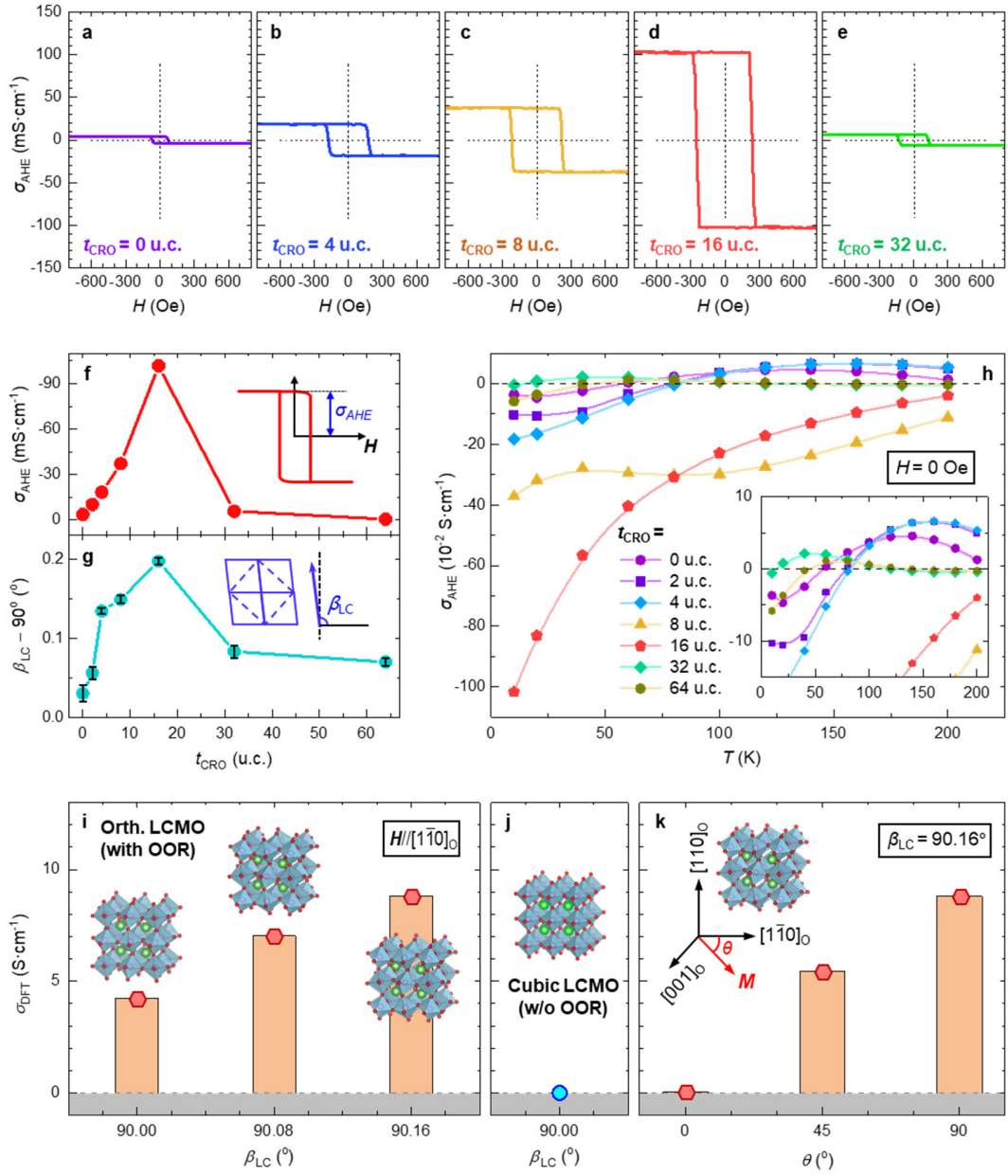

**Fig. 4| Symmetry-tailored IP-AHE of CRO/LCMO/CRO trilayer heterostructures.** (**a-e**) $H$-dependent anomalous Hall conductivity ($\sigma_{AHE}$-$H$) curves measured from a series of CRO/LCMO/CRO/NGO(110)$_O$ heterostructures with various CRO layer thicknesses ($t_{CRO}$). All the measurements are conducted at 10 K with $E$//$H$//LCMO[$\bar{1}$10]$_O$. (**f,g**) $t_{CRO}$-dependent $\sigma_{AHE}$ (f) and 90°-$\beta_{LC}$ (g) values at 10 K. The insets illustrate the definitions of these two parameters. (**h**) Temperature-dependent $\sigma_{AHE}$ ($\sigma_{AHE}$-$T$) curves of CRO/LCMO/CRO/NGO(110)$_O$ heterostructures with various $t_{CRO}$. The inset is a zoom-in plot of $\sigma_{AHE}$-$T$ curves for highlighting the sign changes of $\sigma_{AHE}$. (**i,j**) DFT-calculated $\sigma_{AHE}$ ($\sigma_{DFT}$) of orthorhombic LCMO supercells with various $\beta_{LC}$ values and cubic LCMO supercell ($\beta_{LC}$ = 90°) without any octahedral rotation/distortion. (**k**) Evolution of $\sigma_{DFT}$ with $\theta$ (the angle between in-plane $M$ and the [$\bar{1}$10]$_O$ axis).

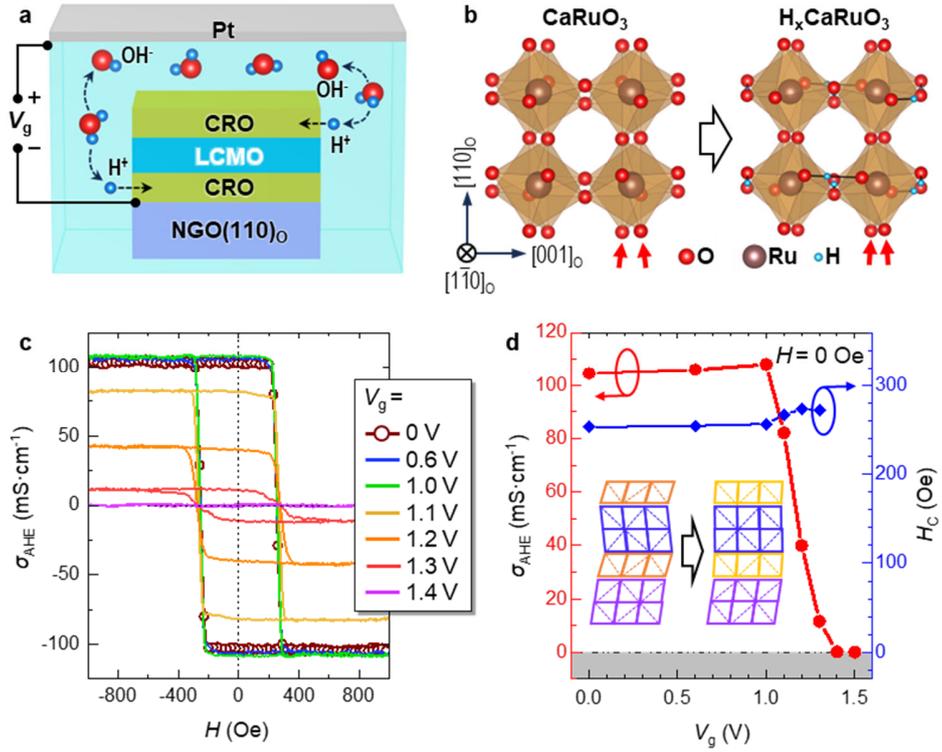

**Fig. 5| Ionic liquid gating of IP-AHE in the CRO/LCMO/CRO trilayer heterostructure.** (**a**) Schematic illustration of the protonation of CRO/LCMO/CRO trilayer through ionic liquid gating. The gating voltage ($V_g$) is applied between the Pt top electrode and the conductive trilayer sample. (**b**) DFT-optimized structure of pristine CaRuO$_3$ and proton-intercalated H$_x$CaRuO$_3$ ($x$ = 1) unit-cells viewed along [$\bar{1}$10]$_O$ axis. The protonation-induced reduction of octahedral tilting is marked by solid (red) arrows. The A-site Ca cations are hidden for clarity. (**c**) $\sigma_{AHE}$-$H$ curves measured from a CRO(16 u.c.)/LCMO(32 u.c.)/CRO(16 u.c.)/NGO(110)$_O$ heterostructure with various $V_g$. All the Hall measurements are conducted at 10 K with ***E***//***H***//LCMO[$\bar{1}$10]$_O$. (**d**) $V_g$-dependent $\sigma_{AHE}$ at zero field and the $H_C$ values extracted from the $\sigma_{AHE}$-$H$ curves. The insets illustrate the reduction of monoclinic-like distortions in CRO and LCMO unit-cells after protonation.